\documentclass[twoside]{ilcws08}
\usepackage[latin1]{inputenc}
\usepackage[dvips]{graphicx,epsfig,color}
\usepackage{wrapfig,rotating}
\usepackage{amssymb,amsmath,array}

\begin{document}

\title{How accurately is the property of the Little Higgs \\ Dark Matter determined at the ILC?}
\author{
Shigeki Matsumoto$^1$, Eri Asakawa$^2$, Masaki Asano$^3$, Keisuke Fujii$^4$, \\
Tomonori Kusano$^5$, Rei Sasaki$^5$, Yosuke Takubo$^5$ and Hitoshi Yamamoto$^4$
\vspace{.3cm}\\
1- Department of Physics, University of Toyama, Toyama, Japan \\
2- Institute of Physics, Meiji Gakuin University, Yokohama, Japan \\
3- Institute for Cosmic Ray Research (ICRR), University of Tokyo, Kashiwa, Japan \\
4- High Energy Accelerator Research Organization (KEK), Tsukuba, Japan \\
5- Department of Physics, Tohoku University, Sendai, Japan
}

\maketitle

\begin{abstract}
Little Higgs scenario provides us a new solution for the hierarchy problem of the Standard Model. The Littlest Higgs model with T-parity, which is one of models in the scenario, solves the dark matter problem in our universe. In this talk, we discuss how accurately the property of the dark matter is determined at linear collides. We also mention how precisely does the thermal abundance of the dark matter determined at the 500 GeV and 1 TeV collides, and compare the results with those at the LHC \cite{Slide}.
\end{abstract}

\section{Introduction}

The Higgs sector in the Standard Model (SM) receives large quadratic mass corrections from top and gauge boson loop diagrams. New symmetries involving top-Higgs and gauge-Higgs sectors below 1 TeV are proposed to remove the corrections. Important new physics candidate is the Minimal Supersymmetric Standard Model. Recently, alternative scenarios which do not rely on Supersymmetry have been proposed. The little Higgs model \cite{Little Higgs} is one of these alternatives. In this model, the Higgs boson is regarded as a pseudo Nambu-Goldstone boson, which originates in the spontaneous breaking of a global symmetry at certain high scale, and the global symmetry protects the Higgs mass from the quadratic corrections.

The simplest version of the model is called the littlest Higgs model \cite{Arkani-Hamed:2002qy}. The global symmetry of the model is SU(5), which is spontaneously broken into SO(5). The part of the SU(5) symmetry is gauged, and the gauge symmetry is SU(2)$_1$$\times$SU(2)$_2$$\times$U(1)$_1$$\times$U(1)$_2$. The top sector is also extended to respect the part of the global symmetry. The model thus contains several new particles such as heavy gauge bosons and top partners. This model, however, predicts a large correction to the electroweak (EW) observables because of the direct mixing between heavy and light gauge bosons after the EW symmetry breaking \cite{Perelstein:2005ka}. The precision EW measurements force the masses of heavy gauge bosons and top partners to be 10 TeV, reintroducing the fine tuning problem to the Higgs mass. A solution of the problem is the introduction of T-parity to the model which forbids the mixing \cite{T-parity}. This is the symmetry under the transformations, SU(2)$_1$$\leftrightarrow$SU(2)$_2$ and U(1)$_1$$\leftrightarrow$U(1)$_2$. Almost all new particles are assigned a T-odd charge, while the SM particles have a T-even charge. The matter sector should be extended so that T-odd partners are predicted. The lightest T-odd particle is the heavy photon $A_H$, which is stable and a candidate for dark matter \cite{LHDM}.

In this talk, we consider the Littlest Higgs model with T-parity (LHT) as a typical example of models implementing both the Little Higgs mechanism and T-parity. We focus mainly on how accurately the property of $A_H$ (dark matter) is determined at the LHC and ILC, and discuss the connection between these collider experiments and cosmology.

\section{Collider Signatures}

We consider production processes of new heavy particles as a collider signature of the LHT; productions of top partners at the Large Hadron Collider (LHC), and those of heavy gauge bosons at the ILC with 500 GeV and 1 TeV center of mass energies. This is because top partners and heavy gauge bosons play an important role in the cancellation of quadratically divergent corrections to the Higgs mass, namely the Little Higgs mechanism. At the LHC, top partners are expected to be copiously produced, and their properties will be determined accurately. However, it is difficult to determine the properties of heavy gauge bosons at the LHC, because they have no color charge. On the other hand, the International Linear Collider (ILC) will provide an ideal environment to measure the properties of heavy gauge bosons. Detections of the signals at both LHC and ILC experiments are therefore mandatory to search the LHT, which leads to the deep understanding of the Little Higgs mechanism.

\begin{wraptable}{l}{0.67\columnwidth}
 \centerline{\small
  \begin{tabular}{|c|c|c|c|}
   \hline
   $f$ (GeV) & $m_h$ (GeV) & $\lambda_2$ & $\kappa_l$ \\
   \hline
   580 & 134 & 1.15 & 0.5 \\
   \hline 
   \hline
   $m_{A_H}$ (GeV) & $m_{W_H(Z_H)}$ (GeV) & $m_{T_+}$ (GeV) & $m_{T_-}$ (GeV) \\
   \hline
   82 & 370 & 840 & 670 \\
   \hline
  \end{tabular}
 }
 \caption{\small Representative point used in our simulation study}
 \label{table:point}
\end{wraptable}
In order to perform a numerical simulation at the collider experiments, we need to choose a representative point in the parameter space of the LHT. We consider the electroweak precision measurements and the WMAP observation to choose a point. Using a well-used $\chi^2$ analysis, we have selected a point shown in Table \ref{table:point}. No fine-tuning is needed at the sample point to keep the Higgs mass on the electroweak scale. The masses of the top partners ($T_+$: T-even top partner, $T_-$: T-odd top partner) and heavy gauge bosons ($A_H$: heavy photon, $W_H (Z_H)$: heavy $W(Z)$ boson) at the point are also summarized in the Table. Notice that, once we fix the parameters $f$ (vacuum expectation value of the breaking; SU(5) $\rightarrow$ SO(5)), $m_h$ (Higgs mass), $\lambda_2$ ($\lambda_2 f$ gives a vector-like mass of the T-odd top partner), and $\kappa_l$ ($\sqrt{2}\kappa_l  f$ gives a vector-like mass of heavy leptons), all masses of new particles and their interactions are completely determined.

We consider following three processes, $pp \rightarrow T_+ \bar{T}_+$, $pp \rightarrow T_+ \bar{t}~(t \bar{T}_+)$, $pp \rightarrow T_- \bar{T}_-$ as signals of the LHT at the LHC. The T-even top partner $T_+$ decays mainly into $t W$, while $T_-$ decays into $t A_H$ with almost 100\% branching fraction. On the other hand, there are several processes for the productions of heavy gauge bosons at the ILC. Among those, we consider the process $e^+e^- \rightarrow A_HZ_H$ for the signal at the ILC with $\sqrt{s} = 500$ GeV (1.9 fb), because $m_{A_H} + m_{Z_H}$ is less than 500 GeV. On the other hand, we focus on $e^+e^- \rightarrow W_H^+W_H^-$ at $\sqrt{s} = 1$ TeV, because huge cross section (280 fb) is expected for the process. Note that $Z_H$ decays into $A_H h$, and $W_H$ decays into $A_H W$ with almost 100\% branching fractions.

\subsection{Signatures at the LHC}

For $T_+ \bar{T_+}$ pair production process, an accurate determination of $m_{T_+}$ is possible. For single-$T_+$ production, we can obtain information on the mixing parameter between $T_+$ and top quark from the measurement of its cross section. For $T_- \bar{T}_-$ pair production, studying the upper end-point of the $M_{T2}$ distribution, a certain relation between $m_{A_H}$ and $m_{T_-}$ is obtained. See Ref.\cite{LHT at LHC} for more details of these simulation studies. Since the top sector in the LHT is parameterized by $f$ and $\lambda_2$, each measurement of these three observables provides a relation between these two parameters. The measurements of the three observables then give non-trivial determinations of the parameters $f$ and $\lambda_2$. It turns out that the parameter $f$ is determined to be \underline{$f = 580 \pm 33$ GeV} at the representative point.

\subsection{Signatures at the ILC ($\sqrt{s} = 500$ GeV)}

\begin{wrapfigure}{r}{0.5\columnwidth}
 \centerline{\includegraphics[width=0.4\columnwidth]{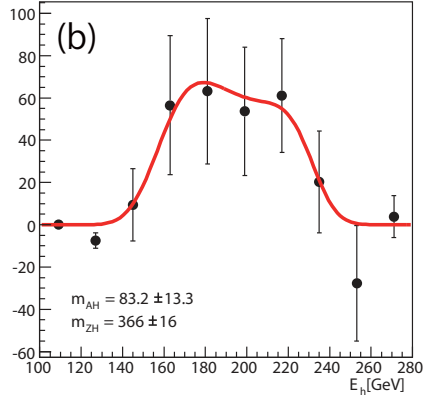}}
 \caption{\small Energy distribution of $h$}
 \label{fig:ahzh}
\end{wrapfigure}
The signature of the $A_HZ_H$ production is a single Higgs boson in the final state, mainly 2 jets from $h \to b \bar{b}$. Importantly, $A_H$ and $Z_H$ masses can be estimated from the edges of the distribution of the reconstructed Higgs boson energies. The signal distribution after backgrounds have been subtracted is shown in Fig.\ref{fig:ahzh}. The endpoints, $E_{\rm max}$ and $E_{\min}$, have been estimated by fitting the distribution with a line shape determined by a high statistics signal sample. The fit resulted in $m_{A_H}$ and $m_{Z_H}$ being $83.2 \pm 13.3$ GeV and $366.0 \pm 16.0$ GeV, respectively. With these values, it turns out that the parameter $f$ is determined to be \underline{$f = 576 \pm 25.0$ GeV}. See Ref.\cite{Asakawa:2009qb} for more details of the simulation study.

\subsection{Signatures at the ILC ($\sqrt{s} = 1$ TeV)}

\begin{wrapfigure}{r}{0.5\columnwidth}
 \centerline{\includegraphics[width=0.4\columnwidth]{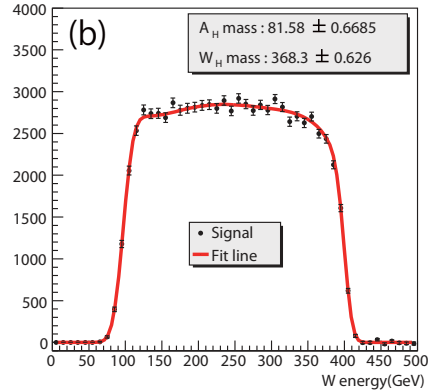}}
 \caption{\small Energy distribution of $W$}
 \label{fig:whwh}
\end{wrapfigure}
Since $W_H$ decays into $A_H$ and $W$ with the 100\% branching fraction, analysis procedure depends on the $W$ decay modes. We have used 4-jet final states from hadronic decays of two $W$ bosons. As in the case of the $A_H Z_H$ production, the masses of $A_H$ and $W_H$ bosons can be determined from the edges of the $W$ energy distribution. The energy distribution of the reconstructed $W$ bosons after subtracting SM backgrounds is depicted in Fig.\ref{fig:whwh}, where the distribution has been fitted with a line shape determined by a high statistics signal sample. The fitted masses of $A_H$ and $W_H$ bosons are $81.58 \pm 0.67$ GeV and $368.3 \pm 0.63$ GeV, respectively. The mass resolution improves dramatically at $\sqrt{s} = 1 $ TeV, compared to that at $\sqrt{s} = 500$ GeV. Using the masses obtained by the simulation, the parameter $f$ is determined as \underline{$f = 580 \pm 0.09$ GeV}. It is also possible to determine the spin and SM gauge charges of the $W_H$ boson, and the helicity of $W$ bosons from the decay of $W_H$ at the 1 TeV ILC experiment. See Ref.\cite{Asakawa:2009qb} for more details of the simulation study.

\section{Cosmological connection}

\begin{wrapfigure}{r}{0.5\columnwidth}
 \centerline{\includegraphics[width=0.4\columnwidth]{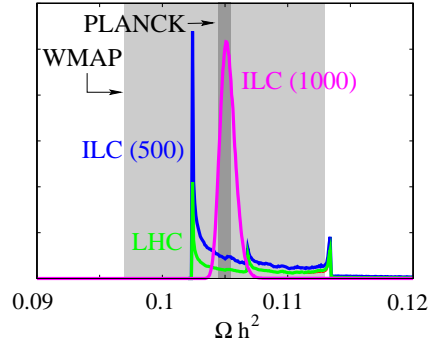}}
 \caption{\small The probability density of $\Omega h^2$}
 \label{fig:Omegah2}
\end{wrapfigure}
Once we obtain the parameter $f$, it is possible to establish a cosmological connection. The most important physical quantity relevant to the connection is the thermal abundance of dark matter relics $\Omega h^2$. It is well known that the abundance is determined by the annihilation cross section of dark matter. In the LHT, the cross section is determined by $f$ and $m_h$ in addition to well known gauge couplings. The Higgs mass $m_h$ is expected to be measured very accurately at the ILC experiment, so that it is quite important to measure $f$ accurately to predict the abundance. Figure \ref{fig:Omegah2} shows the probability density of $\Omega h^2$ obtained from the results in the previous section. As shown in the figure, the abundance will be determined with ${\cal O}$(10\%) accuracy even at the LHC and the ILC with $\sqrt{s} =$ 500 GeV, which is comparable to the WMAP observation \cite{Komatsu:2008hk}. At the ILC with $\sqrt{s} =$ 1 TeV, the accuracy will improve to 1\% level, which stands up to that expected for future cosmological observations such as from the PLANCK satellite \cite{Planck:2006uk}. Note that the results at the ILC are obtained model-independent way, while those at the LHC are obtained using the relation predicted by the model.

\section{Acknowledgments}

The authors thank all the members of the ILC physics subgroup \cite{Ref:subgroup} for useful discussions. They are grateful to the Minami-tateya group for the help of the event generator preparation. This work is supported in part by the Creative Scientific Research Grant (No. 18GS0202) of the Japan Society for Promotion of Science and the JSPS Core University Program.

\begin{footnotesize}

\end{footnotesize}


\begin{thebibliography}{99}

\bibitem{Slide}
  \verb$http://ilcagenda.linearcollider.org/contributionDisplay.py?contribId=192&sessionId=20&confId=2628$

\bibitem{Little Higgs}
    N.~Arkani-Hamed, A.~G.~Cohen and H.~Georgi,
    Phys.\ Lett.\ B {\bf 513} (2001) 232;
    N.~Arkani-Hamed, A.~G.~Cohen, E.~Katz, A.~E.~Nelson, T.~Gregoire and J.~G.~Wacker,
    JHEP {\bf 0208} (2002) 021.

\bibitem{Arkani-Hamed:2002qy}
  N.~Arkani-Hamed, A.~G.~Cohen, E.~Katz and A.~E.~Nelson,
  JHEP {\bf 0207} (2002) 034.

\bibitem{Perelstein:2005ka}
  M.~Perelstein,
  Prog.\ Part.\ Nucl.\ Phys.\  {\bf 58} (2007) 247,
  and references therein.

\bibitem{T-parity}
    H.~C.~Cheng and I.~Low,
    JHEP {\bf 0309} (2003) 051,
    {\bf 0408} (2004) 061;
    I.~Low,
    JHEP {\bf 0410} (2004) 067.

\bibitem{LHDM}
    J.~Hubisz and P.~Meade,
    Phys.\ Rev.\ D {\bf 71} (2005) 035016;
    M.~Asano, S.~Matsumoto, N.~Okada and Y.~Okada,
    Phys.\ Rev.\  D {\bf 75} (2007) 063506;
    A.~Birkedal, A.~Noble, M.~Perelstein and A.~Spray,
    Phys.\ Rev.\  D {\bf 74} (2006) 035002;
    M.~Perelstein and A.~Spray,
    Phys.\ Rev.\  D {\bf 75} (2007) 083519.

\bibitem{LHT at LHC}
    S.~Matsumoto, M.~M.~Nojiri and D.~Nomura,
    Phys.\ Rev.\  D {\bf 75} (2007) 055006;
    S.~Matsumoto, T.~Moroi and K.~Tobe,
    Phys.\ Rev.\  D {\bf 78} (2008) 055018.

\bibitem{Asakawa:2009qb}
  E.~Asakawa {\it et al.},
  arXiv:0901.1081 [hep-ph].

\bibitem{Komatsu:2008hk}
  E.~Komatsu {\it et al.}  [WMAP Collaboration],
  arXiv:0803.0547 [astro-ph].

\bibitem{Planck:2006uk}
  [Planck Collaboration],
  arXiv:astro-ph/0604069.

\bibitem {Ref:subgroup}
  http://www-jlc.kek.jp/subg/physics/ilcphys/.

\end{thebibliography}
\end{document}